\documentclass[%
 reprint,
superscriptaddress,
 longbibliography,
 amsmath,amssymb,
 aps,
prl,
floatfix,
]{revtex4-2}

\usepackage[english]{babel}
\babeladjust{bcp47.toname = on}
\usepackage{graphicx}
\usepackage{dcolumn}
\usepackage{bm,bbm}
\usepackage{xcolor}

\begin{document}

\title{Quantum Mechanism of Piezomagnetism in Higher-Spin Altermagnets}
\author{Daisuke Yamamoto}
\email{yamamoto.daisuke21@nihon-u.ac.jp}
\affiliation{Department of Physics, College of Humanities and Sciences, Nihon University, Sakurajosui, Setagaya, Tokyo 156-8550, Japan}
\affiliation{RIKEN Center for Quantum Computing (RQC), Wako, Saitama 351-0198, Japan}
\affiliation{Global R\&D Center for Business by Quantum-AI Technology (G-QuAT), National Institute of Advanced Industrial Science and Technology (AIST), Tsukuba, Ibaraki 305-8568, Japan}
\author{Makoto Naka}
\email{m-naka@mail.dendai.ac.jp}
\affiliation{Department of Science and Engineering, School of Science and Engineering, Tokyo Denki University, Ishizaka, Saitama 350-0394, Japan}
\date{\today}
\begin{abstract}
{We investigate piezomagnetism in higher-spin altermagnets with easy-plane single-ion anisotropy using a flavor-wave approach. We show that quantum fluctuations of higher-spin collective modes provide a microscopic origin of the piezomagnetic response. For integer spins, the relevant branch softens and evolves into a Higgs-like amplitude mode on approaching the large-$D$ transition, endowing the excitation with a sizable dipolar component and producing a pronounced enhancement of piezomagnetism. By contrast, in half-integer systems the higher-spin branches are progressively separated from the low-energy dipolar sector as the anisotropy increases, which suppresses their contribution to the response. This integer-half-integer contrast directly links macroscopic piezomagnetism to the low-energy fate of multipolar excitations. Our results establish piezomagnetism as a probe of higher-spin quantum dynamics and identify higher-spin altermagnets as a promising setting for quantum magnetoelastic responses.}
\end{abstract}
\maketitle

{\it Introduction.}--- Collective excitations are among the most fundamental manifestations of many-body quantum systems. Beyond serving as fingerprints of ordered phases, they often play an active role in determining macroscopic properties and functionalities~{\cite{Rodin2020}}. Prominent examples include magnons governing spin transport~{\cite{Chumak2015,Pirro2021}}, phonons mediating thermal and structural responses~{\cite{Gu2018}}, and excitons controlling optical properties~{\cite{Wang2018}}. Among the various collective excitations, amplitude (Higgs) modes are particularly intriguing because they correspond to fluctuations in the magnitude of an order parameter and often emerge in the vicinity of {quantum critical points~\cite{Pekker2015,Matsunaga2013-ib,Endres2012,Regg2008}}. Quantum magnets with spin larger than $1/2$ provide a natural platform for such physics, as their enlarged local Hilbert spaces support multipolar degrees of freedom and collective modes beyond conventional dipolar magnons, including quadrupolar and amplitude excitations~{\cite{Penc2010,Romhanyi2012,Bai2021,Jain2017}}. Understanding how these higher-spin collective modes influence measurable material responses remains an important open problem.

Recently, altermagnetism has emerged as a distinct class of collinear magnetic order characterized by momentum-dependent spin splitting despite vanishing net magnetization~{\cite{Ahn2019,Naka2019,Hayami2019,Yuan2020,mejkal2022,Bose2024}}. Owing to this unconventional symmetry property, altermagnets have attracted considerable attention as a promising platform for spintronic functionalities {\cite{Song2025-kk,Jungwirth2026-mn}}. While initial studies focused primarily on electronic band structures, recent works have shown that the altermagnetic symmetry can also manifest itself in collective excitations, giving rise to momentum-dependent spin splitting in magnon spectra {\cite{Naka2019,Smejkal2023-ym,Liu2024-nl}}. These developments have broadened the scope of altermagnetism from electronic phenomena to collective many-body dynamics. However, most studies so far have focused on dipolar excitations, and the role of higher-spin collective modes in altermagnetic systems remains largely unexplored.

A particularly interesting consequence of altermagnetic symmetry is its interplay with magnetoelastic responses. Among them, piezomagnetism, namely the generation of magnetization by lattice distortion, provides a direct route for controlling magnetic states through strain~{\cite{Dzialoshinskii1958,Moriya1959,BorovikRomanov1960,Aoyama2024-dj}}. Recent studies have demonstrated nonrelativistic piezomagnetic effects in altermagnets originating from strain-induced modifications of magnetic exchange interactions~\cite{Yershov2024-kk,Naka2025-bx}. {These studies have so far focused on classical-spin~\cite{Yershov2024-kk} or effective spin-$1/2$~\cite{Naka2025-bx} systems, where the response arises from an occupation imbalance of spin-split excitations through thermal activation~\cite{Yershov2024-kk,Naka2025-bx} or carrier doping~\cite{Naka2025-bx}. It therefore remains unknown whether collective excitations can mediate piezomagnetism through quantum processes at zero temperature and, in particular, what role higher-spin quadrupolar or amplitude modes may play.}

In this Letter, we {uncover such a quantum mechanism of} piezomagnetism in higher-spin altermagnets with easy-plane single-ion anisotropy. Using a flavor-wave approach, we show that {lattice distortions virtually admix higher-lying collective excitations into the ground state, producing a uniform magnetization without requiring their thermal occupation. A characteristic selection rule singles out uniform amplitude branches of multipolar origin as the active modes}. For integer spins, {the lowest active branch evolves from a predominantly quadrupolar excitation toward a dipolar, Higgs-like amplitude mode near the large-$D$ transition}, strongly enhancing the piezomagnetic response. By contrast, for half-integer spins, the corresponding higher-spin excitations are pushed away from the low-energy sector with increasing anisotropy, {preventing a comparable enhancement. Our results establish a distinct quantum route to piezomagnetism and identify higher-spin altermagnets as a platform where otherwise hidden collective modes beyond the conventional low-energy sector become visible through macroscopic magnetic responses.}

\begin{figure}[t]
\includegraphics[scale=0.45]{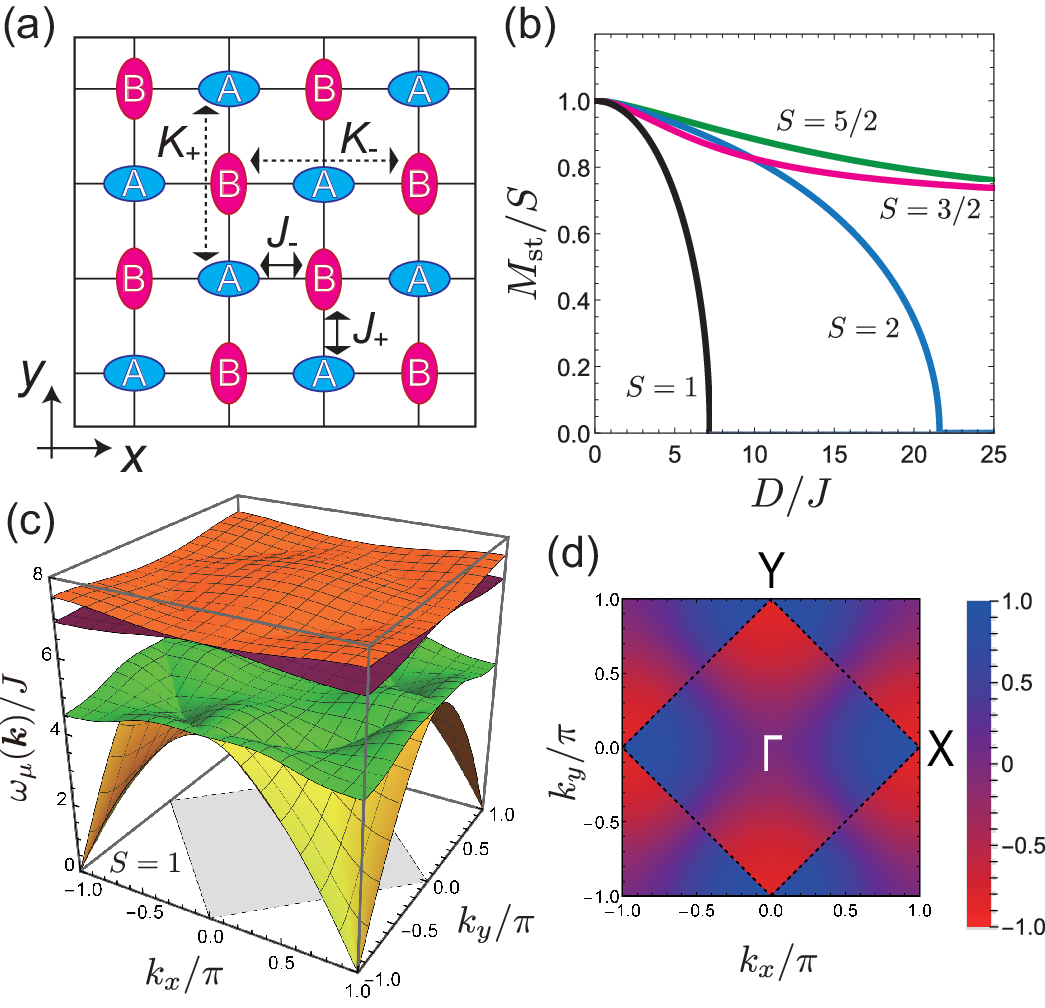}
\caption{\label{fig1}
(a) Square lattice with two sublattices, $\mathrm{A}$ and $\mathrm{B}$, related by a {$C_4$} rotation. In the absence of distortion, the couplings satisfy $J_+=J_-=J$ and $K_+=K_-=K$. Lattice distortions introduce anisotropy as $J_\pm=J\pm\delta J$ and $K_\pm=K\pm\delta K$. (b) Staggered magnetization $M_{\mathrm{st}}/S$ as a function of $D/J$ for $S=1,\,3/2,\,2,$ and $5/2$, at $K/J=0.2$ and in the absence of distortion. (c) Flavor-wave excitation spectrum $\omega_\mu(\bm k)$ for $S=1$ at $K/J=0.2$ and $D/J=2${, plotted in the extended Brillouin zone}. The shaded region at the bottom denotes the first Brillouin zone of the two-site unit cell. (d) Momentum-resolved $m_x(\bm k)$ of the lowest-energy (yellow) excitation in (c).} 
\end{figure}

{\it Higher-spin altermagnetism.}--- We consider a two-sublattice antiferromagnet on a square lattice with spin $S$, where the $\mathrm{A}$ and $\mathrm{B}$ sublattices are related by a {$C_4$} rotation of the lattice in real space, reflecting the underlying altermagnetic symmetry. The system is described by the Hamiltonian
\begin{eqnarray}
\hat{H}_0 &=& J \sum_{\langle i_{\rm A},j_{\rm B} \rangle} \hat{\bm S}_{i_{\rm A}} \cdot \hat{\bm S}_{j_{\rm B}} + K \!\sum_{\langle\!\langle i_{\rm A},j_{\rm A} \rangle\!\rangle_y} \!\hat{\bm S}_{i_{\rm A}} \cdot \hat{\bm S}_{j_{\rm A}}\nonumber
\\ &&
+ K \!\sum_{\langle\!\langle i_{\rm B},j_{\rm B} \rangle\!\rangle_x}\!\hat{\bm S}_{i_{\rm B}} \cdot \hat{\bm S}_{j_{\rm B}}+ D \!\sum_{\eta = {\rm A,B}}\sum_{i_\eta} (\hat{S}_{i_{\rm \eta}}^z)^2,\label{eq:H0}
\end{eqnarray}
where $J$ denotes the nearest-neighbor exchange between the two sublattices, while $K$ represents second-neighbor interactions within each sublattice~{\cite{Naka2019,Naka2025-bx}}, restricted to the $y$ direction for the $\mathrm{A}$ sublattice and to the $x$ direction for the $\mathrm{B}$ sublattice, encoding the sublattice-dependent anisotropy characteristic of altermagnets [see Fig.~\ref{fig1}(a)]. {The last term denotes an easy-plane single-ion anisotropy ($D>0$), relevant for $S\geq1$.}
{For small $D$, t}he ground state of $H_0$ exhibits collinear N\'eel order within the $xy$ plane; we choose the ordered moments to lie along the $x$ {direction. Since the two sublattices are related by $C_4$ rotational symmetry}, the uniform magnetization vanishes. 

Applying a conventional mean-field decoupling to Eq.~(\ref{eq:H0}), we obtain the local mean-field Hamiltonian. Diagonalizing it self-consistently with respect to the staggered magnetization, $M_{\rm st}\equiv\langle \hat{S}_A^x\rangle=-\langle \hat{S}_B^x\rangle$, yields the orthonormal local eigenstates
\begin{equation}
|e_{\mu,\eta}\rangle
=
\sum_{\sigma=-S}^{S}
U^{(\eta)}_{\sigma\mu}
|S^z=\sigma\rangle,
\qquad
\eta=A,B,
\end{equation}
where $\mu=0,\ldots,2S$ labels the eigenstates in ascending energy and $|e_{0,\eta}\rangle$ is the local ground state. Representative results for $M_{\rm st}$ are shown in Fig.~\ref{fig1}(b). For integer spins, increasing the easy-plane anisotropy $D$ drives a transition from the in-plane N\'eel phase ($M_{\rm st}\neq0$) to a large-$D$ quantum paramagnet ($M_{\rm st}=0$), where the local ground state continuously evolves toward $|S^z=0\rangle$. In contrast, no such transition occurs for half-integer spins.

Within flavor-wave theory~{\cite{Papanicolaou1988-ad,Joshi1999-wq,Lauchli2006-hr,Romhanyi2012,Muniz2014-gt}, we introduce $2S$ bosonic flavors $\hat b_{i,\mu}$ ($\mu=1,\dots,2S$) describing local excitations from the mean-field eigenstate $|e_{0,\eta}\rangle$ to $|e_{\mu,\eta}\rangle$}. Introducing the Fourier-transformed fields
\begin{equation}
\hat{\bm b}_{\eta,\bm k}
=
\bigl(
\hat b_{\eta,\bm k,1}~
\hat b_{\eta,\bm k,2}~
\dots~
\hat b_{\eta,\bm k,2S}
\bigr)^{\mathsf T},
\qquad
\eta=\mathrm{A},\mathrm{B},
\end{equation}
we define the Nambu spinor
\begin{equation}
{\hat{\Psi}_{\bm k}^\dagger
=
\left(
\hat{\bm b}_{\mathrm{A},\bm k}^\dagger~
\hat{\bm b}_{\mathrm{B},\bm k}^\dagger~
\hat{\bm b}_{\mathrm{A},-\bm k}^{\mathsf T}~
\hat{\bm b}_{\mathrm{B},-\bm k}^{\mathsf T}
\right).}
\end{equation}
The quadratic Hamiltonian takes the bosonic Bogoliubov--de Gennes form
\begin{equation}
\hat H_0^{(2)}
=
\frac12
\sum_{\bm k}
\hat\Psi_{\bm k}^\dagger
\mathcal H(\bm k)
\hat\Psi_{\bm k},\label{BdG}
\end{equation}
which is diagonalized by a paraunitary Bogoliubov transformation~\cite{Colpa1978-ah},
\(
\hat\Psi_{\bm k}
=
T(\bm k)\hat\Gamma_{\bm k}.
\)
The quasiparticle operators $\hat\Gamma_{\bm k}$ yield $4S$ excitation branches with energies $\omega_\mu(\bm k)$,
\begin{equation}
\hat H_0^{(2)}
=
\sum_{\bm k,\mu}
\omega_\mu(\bm k)
\left(
\hat\gamma_{\bm k,\mu}^\dagger
\hat\gamma_{\bm k,\mu}
+\frac12
\right),
\end{equation}
where $\mu=1,\ldots,4S$ labels the excitation branches.

\begin{figure*}[t]
\includegraphics[scale=0.38]{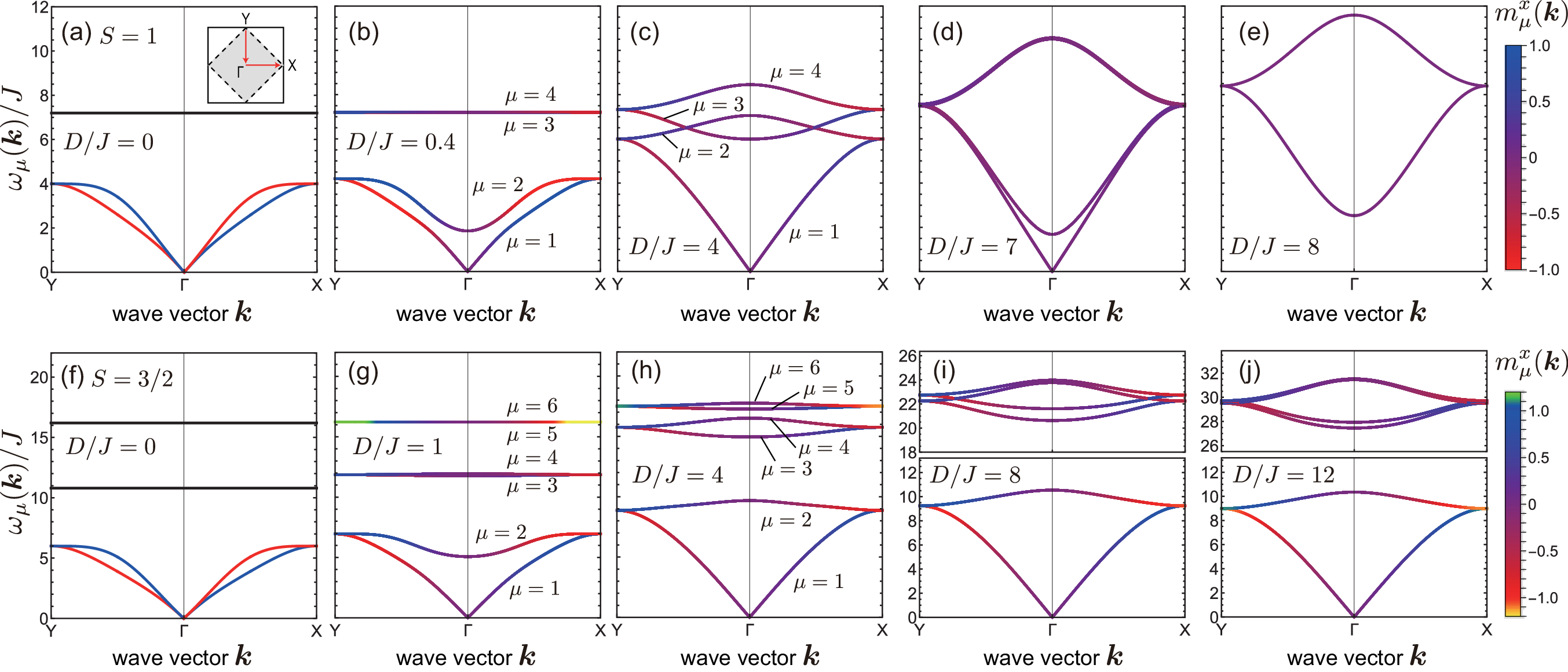}
\caption{\label{fig2}
Evolution of the flavor-wave excitation spectra $\omega_\mu(\bm k)$ with $D/J$ at $K/J=0.2$ along the Y–$\Gamma$–X path in the {first} Brillouin zone, {for (a--e) $S=1$ and (f--j) $S=3/2$. The} color encodes $m_\mu^x(\bm k)$. In (a) and (f), the higher excitation branches at $D/J=0$ are doubly degenerate within the linear flavor-wave approximation, and are shown as black lines due to the ill-defined color. In (i) and (j), the spectra are split into two panels due to the large separation between low- and high-energy branches.
} 
\end{figure*}
%

{F}igure~\ref{fig1}(c) shows a representative spectrum $\omega_\mu(\bm k)$ for $S=1$, consisting of four excitation branches. To characterize the nature of the excitation modes, we define the momentum-resolved uniform magnetization
\begin{equation}
m_\mu^x(\bm k)=\langle\bm k,\mu|\sum_i \hat{S}^x_{i}|\bm k,\mu\rangle-\langle 0|\sum_i \hat{S}^x_{i}| 0\rangle,
\end{equation}
where $|0\rangle$ is the vacuum of the Bogoliubov quasiparticles and $|\bm k,\mu\rangle\equiv \gamma^\dagger_{\bm k,\mu}|0\rangle$. This quantity measures the uniform magnetization carried by each excitation mode and exhibits {the} characteristic $d$-wave-like pattern {expected for altermagnetic spin splitting, as shown in Fig.~\ref{fig1}(d)}. 


{The evolution of the excitation spectrum with $D/J$ is shown in Fig.~\ref{fig2}, where the branch index $\mu$ is assigned in ascending order of excitation energy at small $D/J$ and tracked continuously as $D/J$ increases. At $D=0$, the residual $U(1)$ symmetry about the ordered moment separates the branches into distinct dipolar ($\mu=1,2$) and multipolar ($\mu\geq 3$) sectors. Since the easy-plane anisotropy is transverse to the ordered moment, finite $D$ progressively hybridizes these sectors. For integer spins [Figs.~\ref{fig2}(a--e)], a branch of multipolar origin ($\mu=3$) is driven into the low-energy sector and softens upon approaching the large-$D$ transition. In the large-$D$ quantum paramagnet, the spectrum becomes gapped and the altermagnetic spin splitting disappears together with the magnetic order. For half-integer spins [Figs.~\ref{fig2}(f--j)], by contrast, the branches of multipolar origin harden and progressively separate from the low-energy excitations, leaving an effective spin-$1/2$ dipolar sector while preserving the altermagnetic spin splitting.}

{\it Piezomagnetic Response.}---
We now {introduce small} {uniaxial} lattice distortions that break the {sublattice-exchanging $C_4$} symmetry and {examine the response the uniform magnetization $\hat M_{\rm uni}\equiv \frac{1}{N}\sum_i\hat S_i^x$, where $N$ is the total number spins}. As illustrated in Fig.~\ref{fig1}(a), {we parameterize the {distortion-modified} couplings} as $J_\pm = J\pm\delta J$, $K_\pm = K\pm\delta K$~{\cite{Naka2025-bx}}, together with {the} sublattice-dependent single-ion anisotrop{ies}, $D_{\rm A,B}=D\pm\delta D$. 
The Hamiltonian is then decomposed as
\begin{equation}
\hat{H}
=
\hat{H}_0+\delta J \hat{\mathcal{V}}_J+\delta K \hat{\mathcal{V}}_K+\delta D \hat{\mathcal{V}}_D,
\end{equation}
where
\begin{eqnarray}
\hat{\mathcal{V}}_J &=& 
\sum_{\langle i_{\rm A}, j_{\rm B}\rangle_y}\!\!
\hat{\bm S}_{i_{\rm A}}\cdot\hat{\bm S}_{j_{\rm B}}
-\sum_{\langle i_{\rm A}, j_{\rm B}\rangle_x}\!\!
\hat{\bm S}_{i_{\rm A}}\cdot\hat{\bm S}_{j_{\rm B}},
\\
\hat{\mathcal{V}}_K &=&
\sum_{\langle\!\langle  i_{\rm A}, j_{\rm A}\rangle\!\rangle_y}\!\!
\hat{\bm S}_{i_{\rm A}}\cdot\hat{\bm S}_{j_{\rm A}}
-\sum_{\langle\!\langle  i_{\rm B}, j_{\rm B}\rangle\!\rangle_x}\!\!
\hat{\bm S}_{i_{\rm B}}\cdot\hat{\bm S}_{j_{\rm B}},
\\
\hat{\mathcal{V}}_D &=&
\sum_{i_{\rm A}}(\hat{S}_{i_{\rm A}}^z)^2
-
\sum_{i_{\rm B}}(\hat{S}_{i_{\rm B}}^z)^2.
\end{eqnarray}

{Because both $\hat M_{\rm uni}$ and the distortion perturbations carry zero crystal momentum, their linear couplings involve only the $\bm k=0$ Bogoliubov modes. Defining
$g_\mu\equiv\sqrt{N}\langle 0|\hat M_{\rm uni}|\bm 0,\mu\rangle$ and $f_{\lambda,\mu}
\equiv\frac{1}{\sqrt{N}}\,\langle 0|\hat{\mathcal V}_\lambda|\bm 0,\mu\rangle,
$
where $\lambda=J, K, D$, first-order perturbation theory gives the piezomagnetic susceptibility with respect to each distortion $\delta\lambda$ as
\begin{equation}
\chi_\lambda
\equiv
\left.
\frac{\partial {\langle\hat{M}_{\rm uni}\rangle}}{\partial(\delta\lambda)}
\right|_{\delta\lambda=0}
=
-2
\sum_\mu
\frac{
\mathrm{Re}
\left[
g_\mu^*f_{\lambda,\mu}
\right]
}
{\omega_\mu(\bm 0)}.\label{kai}
\end{equation}
Technical details of the flavor-wave calculation are provided in the End Matter.
} 

\begin{figure}[t]
\includegraphics[scale=0.48]{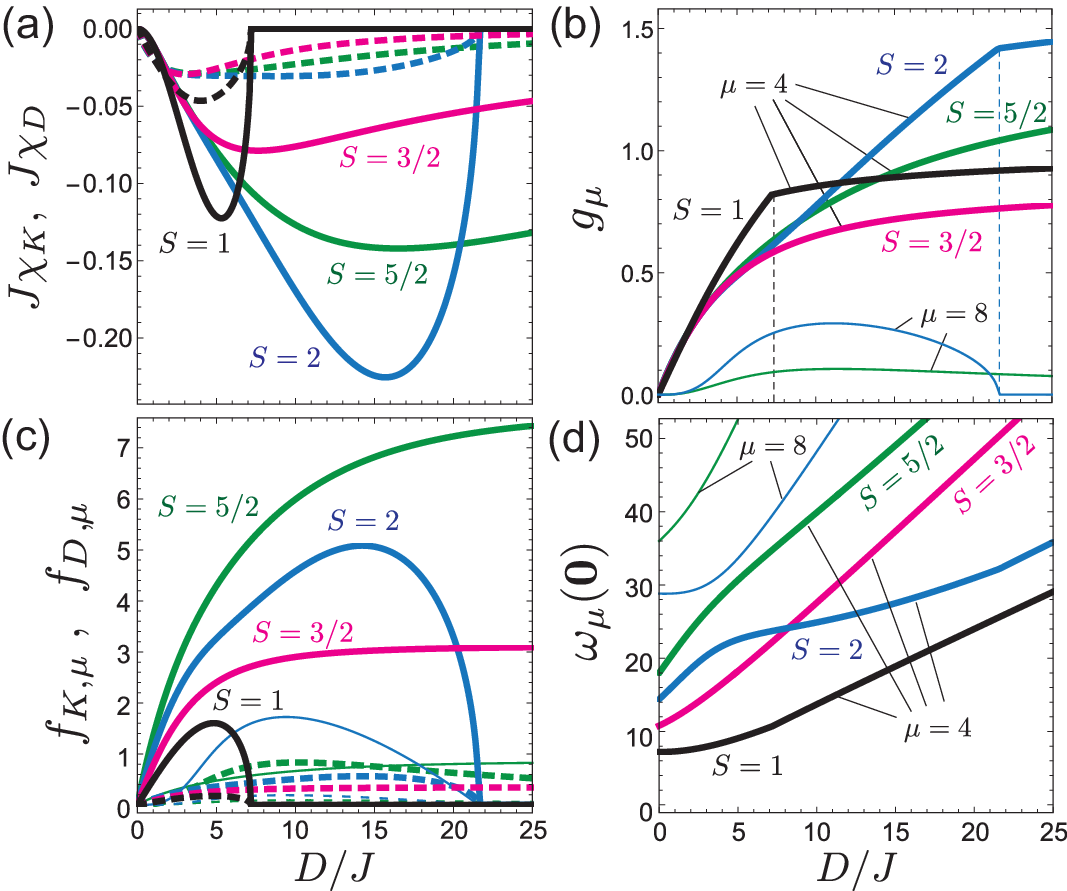}
\caption{\label{fig3}
(a) Piezomagnetic susceptibilities $\chi_K$ (dashed) and $\chi_D$ (solid), multiplied by $J$, as functions of $D/J$ at $K/J=0.2$ for $S=1,\,3/2,\,2,$ and $5/2$. {Mode dependences (a) of $g_\mu$, (b) of $f_{K,\mu}$ (dashed) and $f_{D,\mu}$ (solid), and (c) of $\omega_\mu(\bm{0})$} as functions of $D/J$ for different $S$, with the same color scheme as in (a). {Thick and thin curves in (b)--(d) denote the active branches $\mu=4$ and $8$, respectively; for all other branches, $g_\mu=f_{K,\mu}=f_{D,\mu}=0$}. }
\end{figure}

Within the present linear-response treatment, the nearest-neighbor distortion $\delta J$ does not generate a linear source term at $\bm k=0$, yielding $f_{J,\mu}=0$ and hence $\chi_J=0$. Figure~\ref{fig3}(a) shows the dimensionless susceptibilities $J\chi_K$ and $J\chi_D$ as functions of $D/J$ at fixed $K/J=0.2$ for several integer and half-integer spins, with the $\delta D$ response generally larger in magnitude. For integer spins, the susceptibilities exhibit pronounced nonmonotonic behavior and are strongly enhanced near the transition to the large-$D$ phase. For half-integer spins, the susceptibilities vary more smoothly with $D/J$ and only gradually approach zero as $D/J\to\infty$. {Notably, the peak magnitudes for $S=1$ and $S=2$ exceed those for the larger spins $S=3/2$ and $S=5/2$, respectively.}

{\it Quantum mechanism.}---
Equation~\eqref{kai} reveals the quantum origin of the zero-temperature piezomagnetic response. A lattice distortion virtually admixes the $\bm k=0$ excited states into the ground state through $f_{\lambda,\mu}$, while a finite $g_\mu$ converts this admixture into a uniform magnetization. The response therefore requires no thermal occupation of the excitations. To identify the active modes and elucidate the enhanced integer-spin response, we examine the three ingredients of Eq.~\eqref{kai}, $g_\mu$, $f_{\lambda,\mu}$, and $\omega_\mu(\bm 0)$, shown in Figs.~\ref{fig3}(b)--(d), respectively, with the phase of each $\hat\gamma_{\bm0,\mu}$ chosen such that $g_\mu$ is real and nonnegative.

We first examine the branch dependence of $g_\mu$. Interestingly, $g_\mu$ vanishes identically for all branches except those with $\mu=4n$ ($n=1,2,\ldots$); Fig.~\ref{fig3}(b) therefore shows only these active branches. To elucidate this selection rule, we examine the dipolar spin-fluctuation channels of each mode,
\begin{equation}
\delta S^\alpha_{\pm,\mu}
=\frac{1}{\sqrt N}\left(
\sum_{i_{\rm A}}
\langle 0|\hat S^\alpha_{i_{\rm A}}|\bm 0,\mu\rangle
\pm
\sum_{i_{\rm B}}
\langle 0|\hat S^\alpha_{i_{\rm B}}|\bm 0,\mu\rangle\right),\label{dipole}
\end{equation}
where $\alpha=x,y,z$. As summarized in Table~\ref{tab:modes}, the branches with $\mu=4n-3$ and $4n-2$ are phase modes carrying transverse $(y,z)$ dipolar fluctuations, including the Nambu--Goldstone mode at $\mu=1$, whereas those with $\mu=4n-1$ and $4n$ are amplitude modes carrying staggered and uniform longitudinal $(x)$ fluctuations, respectively. {Although the antiferromagnetic intersublattice coupling places the uniform amplitude branches higher in energy than their staggered counterparts, only the former couple to $\hat M_{\rm uni}$, since $g_\mu=\delta S^x_{+,\mu}$ is nonzero only for $\mu=4n$.} {This selection rule also explains the absence of the present zero-temperature response in classical-spin and $S=1/2$ systems~\cite{Yershov2024-kk,Naka2025-bx}, which lack the relevant uniform amplitude mode.} 

\begin{table}[!t]
\centering
\begin{tabular}{c|c|c|c|c}
$\mu$ & ~mode type~ &~coupled channel(s)~& $g_\mu$ \\
\hline
$4n$ & amplitude & $\delta S^{x}_+$  & ~finite~ \\
$4n-1$ & ~amplitude~ & $\delta S^{x}_-$  & $0$ \\
$4n-2$ & phase & $\delta S^{y}_+$,$\delta S^z_-$  & $0$ \\
~$4n-3$~ & phase & ~{$\delta S^{y}_-$}~  & $0$
\end{tabular}
\caption{
Classification of the $\bm k=0$ excitation branches. The coupled channel(s) list the nonvanishing dipolar matrix elements $\delta S^\alpha_{\pm,\mu}${, where $+$ ($-$) denotes the uniform (staggered) sublattice combination.}}
\label{tab:modes}
\end{table}

The susceptibilities in Fig.~\ref{fig3}(a) are governed primarily by the lowest active branch, $\mu=4$. At small $D$, this mode retains predominantly quadrupolar character and therefore couples only weakly to $\hat M_{\rm uni}$, resulting in a small $g_4$ [Fig.~\ref{fig3}(b)]. Increasing $D$ hybridizes dipolar and quadrupolar fluctuations and enhances $g_4$. In integer-spin systems, the mode further evolves toward a predominantly dipolar, Higgs-like amplitude excitation near the large-$D$ transition, leading to a pronounced enhancement of its coupling to the uniform magnetization.

The distortion matrix elements $f_{K,4}$ and $f_{D,4}$ vanish at $D=0$, where the corresponding perturbations are orthogonal to the active mode, and grow with increasing $D$ [Fig.~\ref{fig3}(c)]. Their growth is stronger for larger $S$, with $|f_{D,4}|$ generally exceeding $|f_{K,4}|$. For integer spins, however, both matrix elements eventually vanish at the large-$D$ transition, as the magnetic order disappears and the two sublattices become equivalent, yielding their characteristic dome-shaped dependence. 

The excitation energy exhibits a qualitatively different integer--half-integer contrast [Fig.~\ref{fig3}(d)]. For integer spins, the $\mu=4$ branch remains at relatively low energy as the uniform partner of the staggered $\mu=3$ amplitude mode that softens toward the transition~{[see also Figs.~\ref{fig2}(a)--(e)]}. For half-integer spins, the corresponding branch is instead rapidly pushed to higher energies with increasing $D$  {[Figs.~\ref{fig2}(f)--(j)]}.

Thus, the pronounced nonmonotonic peaks of $J\chi_K$ and $J\chi_D$ for integer spins in Fig.~\ref{fig3}(a) result from the cooperative action of three effects: the Higgs-like evolution of the active mode enhances $g_4$, its relatively low excitation energy amplifies the virtual admixture through $1/\omega_4(\bm 0)$, and the dome-shaped $f_{\lambda,4}$ determines the peak structure. By contrast, the rapid hardening of the active mode in half-integer systems prevents a comparable enhancement. This mechanism explains why the integer-spin peak susceptibilities exceed the naive trend expected with increasing $S$.

\begin{figure}[t]
\includegraphics[scale=0.48]{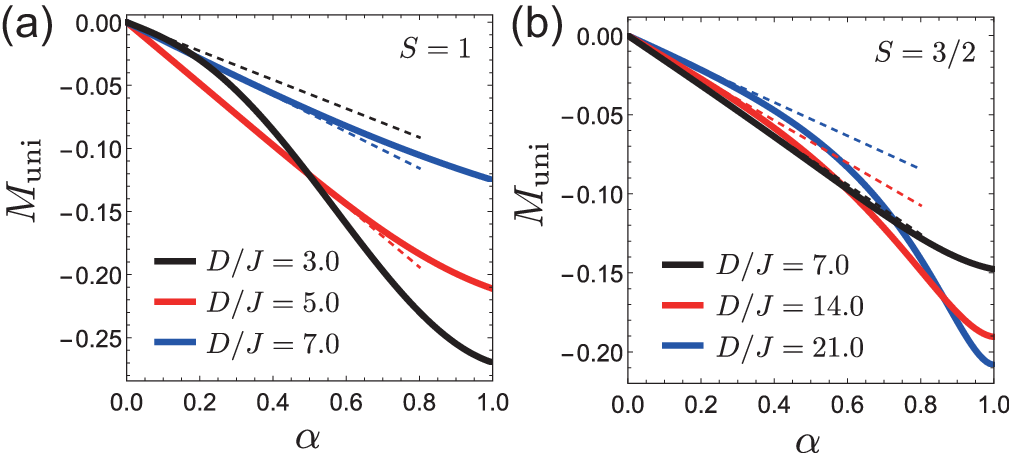}
\caption{\label{fig4}
Strain-induced uniform magnetization $M_{\rm uni}$ as a function of the distortion parameter $\alpha$, obtained from multicomponent mean-field theory with $\delta J=\alpha J$, $\delta K=\alpha K$, and $\delta D=\alpha D$. (a) $S=1$ and (b) $S=3/2$ for different $D/J$ at $K/J=0.2$. Dashed lines show the linear-response prediction $(K\chi_K+D\chi_D)\alpha$.
}
\end{figure}

We also calculate ${M}_{\rm uni}\equiv\langle\hat{M}_{\rm uni}\rangle$ for finite distortions, assuming
$\delta J/J=\delta K/K=\delta D/D=\alpha$.
Figure~\ref{fig4} shows the induced uniform magnetization $M_{\rm uni}$ together with the linear-response prediction
$M_{\rm uni}=(K\chi_K+D\chi_D)\alpha$, where $\chi_J=0$.
The agreement at small $\alpha$ confirms the susceptibility analysis, while nonlinear deviations appear at larger distortions.

{\it Conclusion.}---
We have established a zero-temperature quantum mechanism of piezomagnetism mediated by higher-spin collective excitations. Lattice distortions virtually admix higher-lying modes into the ground state, with a characteristic mod-$4$ selection rule selecting uniform amplitude branches of multipolar origin. For integer spins, the lowest active branch develops dipolar, Higgs-like amplitude character near the large-$D$ transition while remaining at relatively low energy, producing a pronounced enhancement of the response. For half-integer spins, the corresponding branch is instead driven to higher energies, preventing a comparable enhancement. {Among higher-spin altermagnets, easy-plane materials such as
rutile-type NiF$_2$ ($S=1$)~\cite{Hutchings1970,VanHaren2023} and
NiAs-type $\alpha$-MnTe ($S=5/2$)~\cite{Aoyama2024-dj,Liu2024-dx}
provide concrete material settings for the integer- and
half-integer-spin cases discussed above.}

{Although demonstrated in a minimal local-moment model, the mechanism relies more generally on the virtual mixing of distortion-active collective modes carrying uniform-magnetization weight and should apply to systems with analogous amplitude or multipolar excitations. Crucially, it does not require altermagnetic spin splitting: even at $K=0$, a sublattice-dependent modulation $\delta D$ induces a finite zero-temperature response, distinguishing it from mechanisms based on an occupation imbalance of spin-split excitations~\cite{Yershov2024-kk,Naka2025-bx}. More generally, our findings show that higher-lying collective excitations can control static magnetoelastic responses through virtual quantum processes.}

\begin{acknowledgments}
{The work was supported by JSPS KAKENHI Grant Numbers~ JP23K25830 (DY), JP24K06890 (DY), JP26K00664 (DY), JP23K25826 (MN), JP23K03333 (MN), JP25H00838 (MN), and JST PRESTO Grant No.~JPMJPR245D (DY).}
\end{acknowledgments}

\bibliography{apsbib_cleaned}

\clearpage
\appendix
\onecolumngrid
\begin{center}
{\bfseries END MATTER}
\end{center}
\twocolumngrid

{\it Linear flavor-wave formulation.}---
We provide details of the linear flavor-wave calculation~\cite{Papanicolaou1988-ad,Joshi1999-wq,Lauchli2006-hr,Muniz2014-gt} used in the
main text. For each sublattice $\eta={\rm A,B}$, we diagonalize the
local mean-field Hamiltonian as
\begin{equation}
\hat h_\eta^{\rm MF}|e_{\nu,\eta}\rangle
=
\varepsilon_{\nu,\eta}|e_{\nu,\eta}\rangle,
\qquad
\nu=0,\ldots,2S,
\end{equation}
where the eigenstates are ordered by energy and
$|e_{0,\eta}\rangle$ denotes the local ground state. Introducing one
flavor boson for each local state,
\begin{equation}
\hat b_{i\eta,\nu}^\dagger|{\rm vac}\rangle
=
|e_{\nu,\eta}\rangle,
\qquad
\sum_{\nu=0}^{2S}
\hat b_{i\eta,\nu}^\dagger\hat b_{i\eta,\nu}=1,
\end{equation}
the mean-field state corresponds to condensation of the $\nu=0$
boson. We eliminate this condensed flavor through
\begin{equation}
\hat b_{i\eta,0}
=
\left(
1-\sum_{\nu=1}^{2S}
\hat b_{i\eta,\nu}^\dagger\hat b_{i\eta,\nu}
\right)^{1/2}
\simeq
1-\frac{1}{2}
\sum_{\nu=1}^{2S}
\hat b_{i\eta,\nu}^\dagger\hat b_{i\eta,\nu}.
\end{equation}
The spin operators are then expanded as
\begin{align}
\hat S_{i\eta}^{\alpha}
={}&
S_{\eta,00}^{\alpha}
+
\sum_{\nu=1}^{2S}
\left(
S_{\eta,0\nu}^{\alpha}\hat b_{i\eta,\nu}
+
S_{\eta,\nu0}^{\alpha}\hat b_{i\eta,\nu}^\dagger
\right)
\nonumber\\
&+
\sum_{\nu,\nu'=1}^{2S}
\left(
S_{\eta,\nu\nu'}^{\alpha}
-
S_{\eta,00}^{\alpha}\delta_{\nu\nu'}
\right)
\hat b_{i\eta,\nu}^\dagger
\hat b_{i\eta,\nu'}
+\cdots ,
\label{eq:spin_expansion}
\end{align}
where $S_{\eta,\nu\nu'}^\alpha
\equiv
\langle e_{\nu,\eta}|
\hat S^\alpha
|e_{\nu',\eta}\rangle$.

Using
\begin{equation}
\hat b_{i\eta,\nu}
=
\sqrt{\frac{2}{N}}
\sum_{\bm k}
e^{i\bm k\cdot\bm r_i}
\hat b_{\eta,\bm k,\nu},
\end{equation}
we introduce
$\hat{\bm b}_{\eta,\bm k}
=
\bigl(
\hat b_{\eta,\bm k,1}~
\hat b_{\eta,\bm k,2}~
\dots~
\hat b_{\eta,\bm k,2S}
\bigr)^{\mathsf T}$.
The terms linear in the bosons vanish owing to the mean-field self-consistency condition, and the quadratic Hamiltonian takes the bosonic Bogoliubov--de Gennes form given in Eq.~\eqref{BdG} of the main text, with the $8S\times8S$ Hamiltonian matrix $\mathcal H_{\bm k}$.
It is diagonalized by a paraunitary transformation
$\hat{\bm\Psi}_{\bm k}=\mathcal T_{\bm k}
\hat{\bm\Gamma}_{\bm k}$ satisfying
\begin{equation}
\mathcal T_{\bm k}^\dagger
\Sigma\mathcal T_{\bm k}
=
\Sigma,
\qquad
\Sigma=
\operatorname{diag}
\left(
\mathbbm 1_{4S},-\mathbbm 1_{4S}
\right),
\end{equation}
where $\mathbbm 1_{4S}$ denotes the $4S\times4S$ identity matrix~\cite{Colpa1978-ah}.
This transformation yields $4S$ positive-energy excitation branches
$\omega_\mu(\bm k)$, with $\mu=1,\ldots,4S$.

Since both $\hat M_{\rm uni}$ and $\hat{\mathcal V}_\lambda$ are spatially uniform, their terms linear in the Bogoliubov quasiparticles involve only the $\bm k=\bm0$ modes:
\begin{align}
\hat M_{\rm uni}^{(1)}
&=
\frac{1}{\sqrt N}
\sum_\mu
\left(
g_\mu\hat\gamma_{\bm0,\mu}
+
g_\mu^*\hat\gamma_{\bm0,\mu}^\dagger
\right),
\\
\hat{\mathcal V}_\lambda^{(1)}
&=
\sqrt N
\sum_\mu
\left(
f_{\lambda,\mu}\hat\gamma_{\bm0,\mu}
+
f_{\lambda,\mu}^*
\hat\gamma_{\bm0,\mu}^\dagger
\right).
\end{align}
For a perturbation
$\delta\lambda\,\hat{\mathcal V}_\lambda$, the first-order correction to the ground state is
\begin{equation}
|\delta0\rangle
=
-\delta\lambda\sqrt N
\sum_\mu
\frac{f_{\lambda,\mu}^*}
{\omega_\mu(\bm0)}
|\bm0,\mu\rangle,
\end{equation}
reproducing Eq.~\eqref{kai} of the main text. For the nearest-neighbor perturbation $\hat{\mathcal V}_J$, the contributions from the $x$- and $y$-directed bonds cancel at $\bm k=\bm0$, yielding
$f_{J,\mu}=0$ and hence $\chi_J=0$.

{\it Dipolar and quadrupolar mode contents.}---
To complement the mode classification in Table~\ref{tab:modes}, we provide the
dipolar matrix elements $\delta S^\alpha_{\pm,\mu}$ defined in
Eq.~\eqref{dipole}, together with the quadrupolar
matrix elements
\begin{align}
\delta Q^{\alpha\beta}_{\pm,\mu}
&\equiv
\frac{1}{\sqrt N}
\left[
\sum_{i_{\rm A}}
\langle0|\hat Q^{\alpha\beta}_{i_{\rm A}}|\bm0,\mu\rangle
\pm
\sum_{i_{\rm B}}
\langle0|\hat Q^{\alpha\beta}_{i_{\rm B}}|\bm0,\mu\rangle
\right],
\end{align}
where
\begin{equation}
\hat Q_i^{\alpha\beta}
=
\frac{1}{2}
\left(
\hat S_i^\alpha\hat S_i^\beta+
\hat S_i^\beta\hat S_i^\alpha
\right)
-\frac{\delta_{\alpha\beta}}{3}S(S+1).
\end{equation}

\begin{figure*}[t]
\includegraphics[scale=0.38]{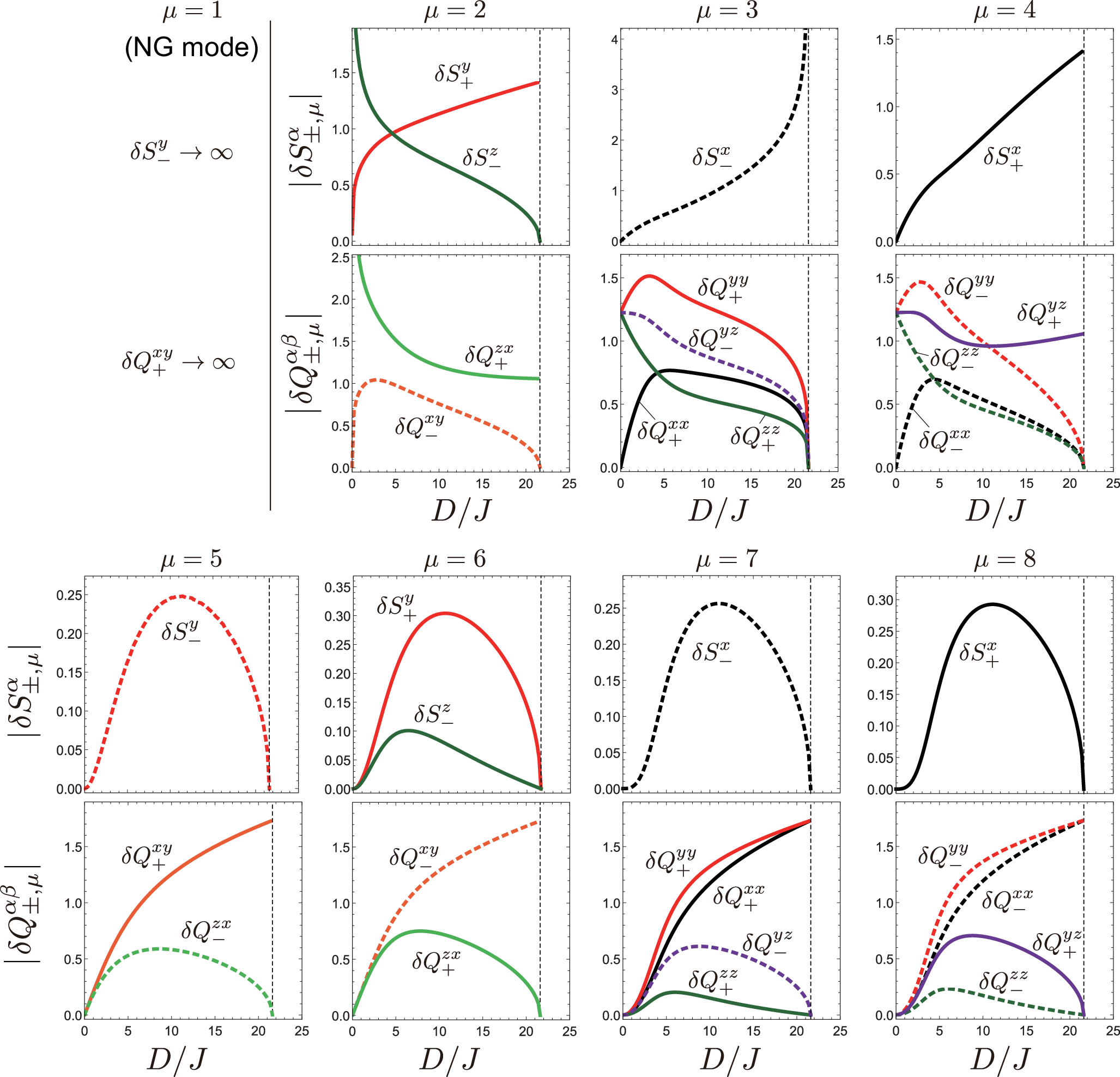}
\caption{\label{fig5}
Absolute values of the nonvanishing normalized dipolar and quadrupolar matrix elements,
$|\delta S^\alpha_{\pm,\mu}|$ and $|\delta Q^{\alpha\beta}_{\pm,\mu}|$,
as functions of $D/J$ for $S=2$ and $K/J=0.2$.
The upper and lower panels show the dipolar and quadrupolar components, respectively, for $\mu=1$--$8$.
For $\mu=1$, only the singular components at $\bm{k}=\bm{0}$ are indicated schematically:
$|\delta S^y_{-,1}|\to\infty$ and $|\delta Q^{xy}_{+,1}|\to\infty$.
The same colors represent the same components throughout, while solid (dashed) lines denote uniform $(+)$ [staggered $(-)$] channels.
Vertical dashed lines indicate the large-$D$ transition, and the matrix elements are shown only in the ordered phase.
The singular behavior of the $\mu=2$ branch as $D\to0$ reflects the restoration of spin-rotation symmetry at $D=0$.
Only nonvanishing components are shown. 
}
\end{figure*}

Figure~\ref{fig5} presents all nonvanishing dipolar and quadrupolar components for $S=2$ and $K/J=0.2$. We choose $S=2$ because its eight branches display two complete repetitions of the mod-$4$ structure and allow the evolution toward the large-$D$ transition to be followed. Half-integer spins exhibit the same qualitative channel structure and anisotropy-induced mixing, except for the absence of the large-$D$ transition. The dipolar components make explicit the mod-$4$ structure summarized in Table~\ref{tab:modes}: the branches $\mu=4n-3$ and $4n-2$ carry transverse fluctuations, while $\mu=4n-1$ and $4n$ carry staggered and uniform longitudinal fluctuations, respectively. In particular, $g_\mu=\delta S^x_{+,\mu}$, so that only the $\mu=4n$ branches couple to the uniform magnetization.

The multipolar origin of these active branches is also evident. At $D=0$, the lowest active branch $\mu=4$ has vanishing dipolar weight, $\delta S^x_{+,4}=0$, while retaining finite quadrupolar components. Finite single-ion anisotropy mixes these sectors and generates the growing uniform longitudinal component $\delta S^x_{+,4}$. For the higher active branch $\mu=8$, both the dipolar and quadrupolar components are identically zero at $D=0$, indicating an excitation of still higher multipolar rank, which likewise acquires lower-rank components at finite $D$.

The Nambu--Goldstone (NG) branch $\mu=1$ is exceptional: its phase-fluctuation components $|\delta S^y_{-,1}|$ and $|\delta Q^{xy}_{+,1}|$ diverge at $\bm k=\bm0$ and are therefore shown only schematically. The singular behavior of the $\mu=2$ branch as $D\to0$ reflects the restoration of spin-rotation symmetry and the emergence of a second Goldstone mode. The curves are restricted to the ordered phase, since degeneracies in the large-$D$ phase render individual mode-resolved matrix elements basis dependent.
\end{document}